# Transition path dynamics of a nanoparticle in a bistable optical trap

Niels Zijlstra[1,4], Daniel Nettels[1], Rohit Satija[2], Dmitrii E. Makarov[2], Benjamin Schuler[1,3]

[1]Department of Biochemistry, University of Zurich, 8057 Zurich, Switzerland.

[2]Department of Chemistry and Oden Institute for Computational Engineering and Sciences, University of Texas at Austin, Austin, Texas 78712, USA.

[3]Department of Physics, University of Zurich, 8057 Zurich, Switzerland.

[4]Current address: Physical and Synthetic Biology, Faculty of Biology, Ludwig-Maximilians-Universität München, Großhadernerstrasse 2-4, 82152 Planegg-Martinsried, Germany

To whom correspondence may be addressed:

N. Zijlstra, n.zijlstra@lmu.de, D. Nettels, nettels@bioc.uzh.ch, D. E. Makarov, makarov@cm.utexas.edu, B. Schuler, schuler@bioc.uzh.ch

**Many processes in chemistry, physics, and biology involve rare events in which the system escapes from a metastable state by surmounting an activation barrier. Examples range from chemical reactions, protein folding, and nucleation events to the catastrophic failure of bridges[1-4]. A challenge in understanding the underlying mechanisms is that the most interesting information is contained within the rare transition paths, the exceedingly short periods when the barrier is crossed[3]. To establish a model process that enables access to all relevant timescales, although highly disparate, we probe the dynamics of single dielectric particles in a bistable optical trap in solution. Precise localization by high-speed tracking enables us to resolve the transition paths and relate them to the detailed properties of the 3D potential within which the particle diffuses. By varying the barrier height and shape, the experiments provide a stringent benchmark of current theories of transition path dynamics.**

Rate processes take place on timescales that are long compared to the relaxation dynamics within the metastable states. This separation of timescales allows the interconversion between states to be approximated in terms of phenomenological chemical kinetics by employing a set of rate coefficients and effectively assuming that the actual interconversion between microscopic states is instantaneous[1]. For microscopic systems, a Kramers-type description of the underlying dynamics is more realistic, where the rate coefficients are related to the diffusive motion on an energy surface[5]. This concept has been particularly successful when applied to reactions in solution, including complex processes such as protein folding and biomolecular recognition[3,5-7]. Such descriptions provide a conceptual link between the phenomenological kinetics and the underlying molecular dynamics, including the actual process of barrier crossing. These transitions across the barriers contain the most interesting information about the mechanism of the reaction[3], and the investigation of transition paths connecting metastable states in biomolecular transformations has made great progress in the past ten years owing to advances in single-molecule experiments and the concomitant development of theoretical concepts[3,4,8-16].

Experimental limitations and the complexity of biomolecular systems, however, pose considerable challenges in data interpretation. For instance, the microsecond timescale of typical biomolecular transition paths seriously limits the amount of information accessible with experimental time resolution; the choice of reaction coordinates is constrained by the experimentally accessible observables; the actual underlying potentials are often unknown; and data analysis thus requires simplifying assumptions regarding the shape of the energy surface and the nature of the dynamics. To complement the ongoing efforts to overcome such difficulties, our goal here is to provide a simple experimental system that enables all relevant parameters to be extracted directly from the measurements; that covers a broad range of timescales to enable both extensive sampling of the rare transitions and the necessary time resolution to fully resolve the transition paths; and that allows the shape of the potential to be tuned systematically. Such a system would enable a stringent comparison to simulations and a rigorous test of theoretical concepts at a level of detail that has previously been inaccessible.

To achieve this goal, we employ a spherical dielectric Brownian particle trapped in an optical double well potential (see Methods & Supplementary Figure 1). This configuration has previously been used to test essential aspects of Kramers theory, such as the calculation of rate coefficients based on the shape of the potential[17], the dependence of rates on the damping regime[18], and the transition between ballistic and diffusive motion[19]. This approach is also ideally suited for the analysis of transition paths, since high-speed camera-based tracking enables 3D localization at microsecond time resolution combined with hour-long recording times[20-22], a prerequisite for extensively sampling rare transitions while simultaneously time-resolving them. Moreover, the laser power can be used to adjust the trapping potential and the barrier height. Finally, all parameters required for analyzing the barrier-crossing dynamics in terms of diffusion in a 3D potential can be extracted directly from the experimental observations: the detailed shape of the potential from Boltzmann inversion of the probability density of the particle's location, and the particle's diffusion coefficient from its dynamics at short times[17].

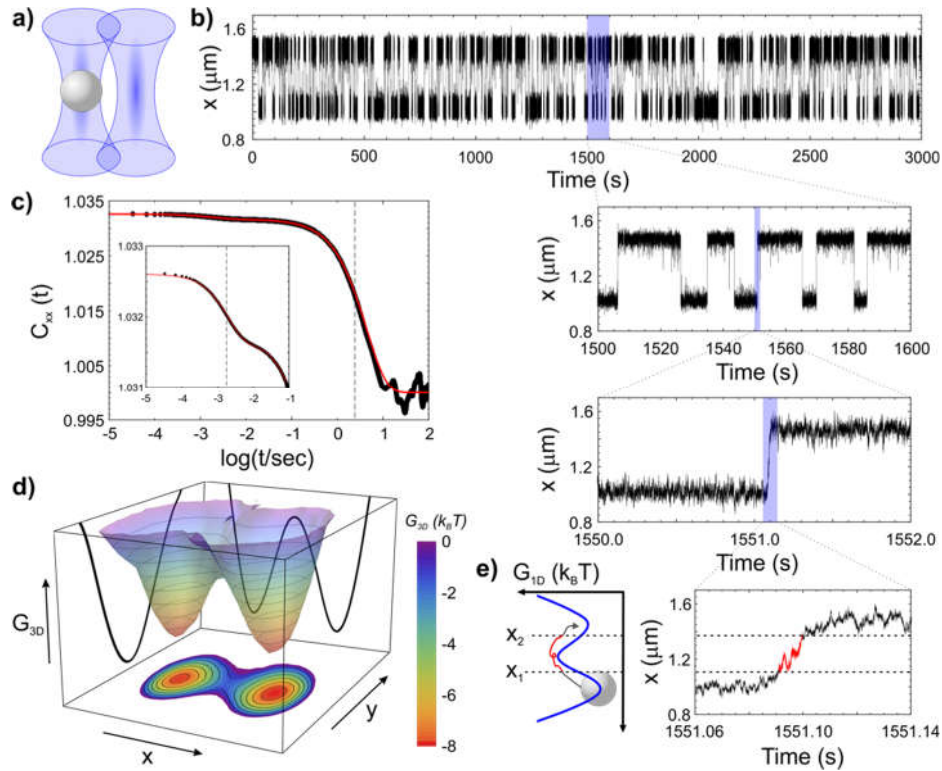

**Figure 1. Transition paths in the bistable optical trap. (a)** Schematic of dielectric particle in a bistable trap formed by two laser beams. **(b)** By imaging the particle with a high-speed camera, its position, $x(t)$, can be followed in 3D with a time resolution of 33 μs for up to about one hour. The trajectory at the highest time resolution (bottom right) illustrates that transition paths can be clearly resolved. **(c)** The position autocorrelation function, $C_{xx}$, shows components corresponding to the diffusive dynamics within the traps (relaxation time of 2.2 ± 0.1 ms in this example, see inset with zoom) and to the kinetics of transitions between them (relaxation time of 4.4 ± 0.1 s in this example, close to the inverse sum of the transition rate coefficients, 3.9 ± 0.3 s). The red line is a fit using $C_{xx}(t) = \langle x \rangle^2 e^{-kt} + (k_B T/\kappa)\, e^{-(D\kappa/k_B T)t} + 1$, where $\langle x \rangle$ is the center between the two wells, $k$ is the transition rate between the wells, $\kappa$ is the average curvature of the free energy at the bottom of the two wells, and $D$ is the diffusion constant. **(d)** The full 3D potential is extracted directly from the distribution of the particle positions, as illustrated by the free energy, $G_{3D}$, shown in the color projection. **(e)** Illustration of a transition path between coordinate boundaries $x_1$ and $x_2$ along the one-dimensional free energy profile, $G_{1D}(x)$. The duration of the transition path (red) in this example is 8.8 ms.

We monitored the motion of individual particles in the bistable optical trap with 33-μs time resolution for up to hours (Figure 1, Supplementary Figure 2), corresponding to a range of eight orders of magnitude in time. The data thus provide both the requisite statistics for obtaining precise transition rates from individual trajectories and at the same time the ability to resolve the millisecond transition paths expected for a particle with a radius of ~270 nm and an inter-focus distance of 450 nm (Figure 1e). Particle positions were determined with a precision of ~50 nm axially and ~25 nm laterally, and their time dependence shows the expected bistability of the system (Figure 1b). The position autocorrelation function (Figure 1c) illustrates the broad range of timescales accessible and reveals both the diffusive dynamics within the traps and the kinetics of transitions between them. From the distribution of the particle's position, $P_{3D}(x,y,z)$, based on $10^6$ to $10^8$ images per trajectory, the 3-dimensional (free) energy landscape is readily reconstructed as $G_{3D}(x,y,z) = -k_B T \ln P_{3D}(x,y,z)$, where $k_B$ is Boltzmann's constant and $T$ is temperature (Figure 1d). The diffusion coefficient, $D$, of each particle was determined from its mean squared displacement at short times with a precision of $\pm 0.01$ µm$^2$s$^{-1}$ (Supplementary Figure 3). Notably, variations in particle shape and size lead to variations of ~10% in $D$ and clearly resolved differences in the potential, including the barrier shape (Supplementary Figures 3 and 4). Each particle must therefore be analyzed individually to account for these differences. By adjusting the laser power, we can tune the barrier heights systematically between ~2 and 8 $k_B T$, corresponding to a hundredfold change in transition rate coefficients (Figure 2). The accessible range of rates is limited at low laser power by the loss of persistent trapping, and at high laser power by insufficient sampling of the potential in the transition region.

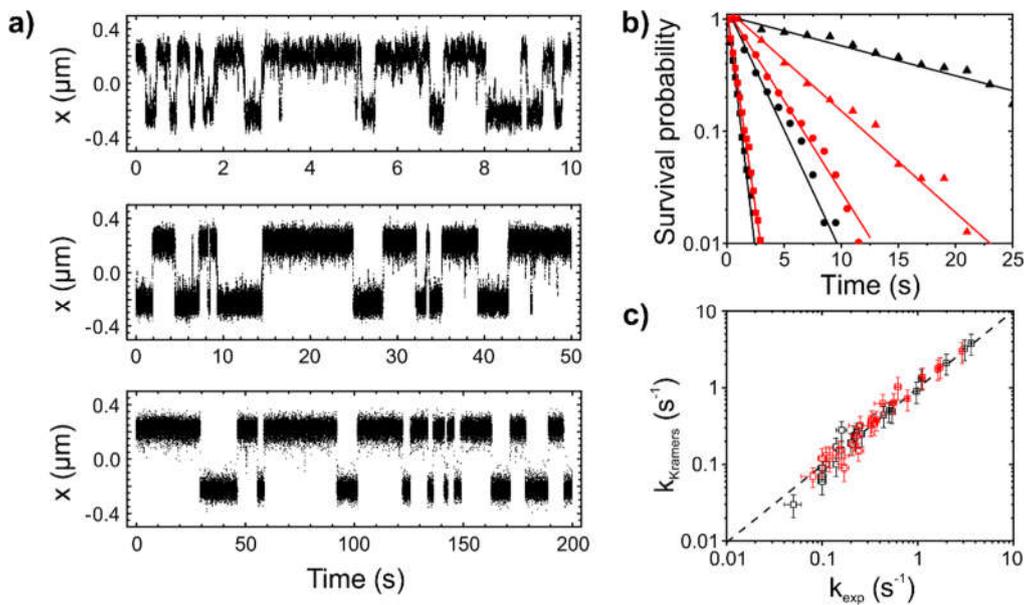

**Figure 2. Transition rates and Kramers theory. (a)** Three examples of particle trajectories projected on the inter-focus axis, *x* (cf. Figure 1a), with very different transition frequencies, resulting from different barrier heights. **(b)** The corresponding survival probabilities in the trap wells (symbols) fitted by exponential decays (lines) yield the inter-well transition rate coefficients. Since the potentials are never perfectly symmetric, transitions in the two directions are analyzed separately (black symbols: *x* increasing; red symbols: *x* decreasing, identical symbols: same particle). **(c)** Comparison of the transition rate coefficients according to Kramers theory, $k_{Kramers}$, with measured values, $k_{exp}$ (black symbols: *x* increasing; red symbols: *x* decreasing; identity line: dashed). To determine $k_{Kramers}$ for each individual particle (Eq. 2), the barrier heights and the curvatures of the potentials at the barrier and in the wells were determined from $G_{1D}(x)$, and the diffusion coefficients were determined from the mean-squared displacement of each particle at short times (Supplementary Figures 3, 4, 5). Error bars for $k_{Kramers}$ result from the uncertainties in the barrier/well curvatures and barrier heights, error bars in $k_{exp}$ from the uncertainties in the fits of the survival probability distributions. Results were obtained from 24 recorded trajectories of different particles and/or laser powers, with a total of 7637 transitions.

As the friction force on the particle obeys Stokes' law and is proportional to its velocity, we expect the dynamics of the particle to be described by a three-dimensional Langevin equation in the potential $G_{3D}(x,y,z)$. Since the particle's motion is overdamped[17,18], inertial contributions can be neglected[5]. The kinetics of transitions between the wells can then be described using Langer's rate theory of multidimensional diffusive barrier crossing[23]. The symmetry of the potential suggests that the 'reaction coordinate' of this process coincides with the *x*-axis along the inter-focus distance (Figure 1); moreover, the dynamics along *x* are decoupled from those of *y* and *z* near the barrier saddle. As a result, Kramers' model[24] of one-dimensional diffusive barrier crossing in the 1D free-energy landscape $G_{1D}(x)= -k_BT \ln P_{1D}(x)$ is sufficient, where $P_{1D}(x) = \iint P_{3D}(x,y,z)\, dy\, dz$ is the probability distribution function of the position along *x*. The Kramers rate for crossing a barrier of height $\Delta G^{\ddagger}$ is $k = D\sqrt{\kappa_b \kappa_w}/2\pi\, e^{-\Delta G^{\ddagger}/k_BT}$, where $\kappa_b$ and $\kappa_w$ are the curvatures of the potential $G_{1D}(x)$ at the top of its barrier and at the bottom of the initial potential well, respectively. Note that *D* is independent of position in our experimental system. To test the applicability of Kramers theory, we extracted transition rate coefficients from the measured particle trajectories projected on *x*, assuming first-order interconversion kinetics between the two metastable states (Figure 2), and compared them to the values from a Kramers model using $G_{1D}(x)$ (Supplementary Figure 5) and *D* measured for the respective individual particles (see Methods). The agreement (Figure 2c) confirms the applicability of Kramers theory[17], illustrates the data quality available from individual particles, and sets the stage for the investigation of transition paths.

Examples of individual transitions from one potential well to the other are shown in Figure 3a, projected on the xy-plane and along x. Transition paths are the segments of the trajectory where the particle continuously dwells in the transition region between the two potential wells, having entered it from one well and exiting to the other (red segments in Figure 3a). We used two different definitions of the boundaries of the transition region: (1) For the trajectories in 3D, we defined them as isoenergetic surfaces enclosing the two potential minima described by $G_{3D}(x,y,z) = E_1$ and $G_{3D}(x,y,z) = E_2$, where $E_1$ and $E_2$ were chosen as the energy values halfway between the barrier top and the bottom of the respective minima of the potential, $G_{3D}(x,\langle y\rangle,\langle z\rangle)$. (2) For the 1D projection, $x(t)$, the boundaries correspond to the values $x_1$ and $x_2$ defined by $G_{1D}(x) = E_1$ and $G_{1D}(x) = E_2$, where the energies $E_1$ and $E_2$ were chosen to be halfway between the barrier top and the respective minima of the potential, $G_{1D}(x)$. The transition region then coincides with the interval $[x_1, x_2]$. We note that the transition paths identified in 3D differ from those identified based on a projection onto one or two coordinates, reflecting the different definitions of the boundaries (Figure 3). Nevertheless, the mean transition path times identified in 1D are on average only ~15% shorter than in 3D, indicating that the results are quite robust to the dimensionality of the analysis used (Supplementary Figure 6).

The average transition path times for each particle and laser power are well reproduced by Brownian dynamics simulations on the corresponding potentials, over a wide range of barrier heights and for different choices of transition path boundaries (Figure 3b). The agreement is similarly good for the numerical solution of the diffusion problem using the detailed shapes of the potentials.[25] Remarkably, the average transition path times, $\langle\tau_{TP}\rangle$, for all but the lowest barriers are well reproduced by the approximation of A. Szabo[3,4,26], according to which $\langle\tau_{TP}\rangle = k_BT/D\kappa_b \ln(2e^{\gamma} \Delta G^*/k_BT)$, where $\Delta G^*$ is the barrier height measured relative to the values of $G_{1D}(x)$ at the transition path boundaries and $\gamma \approx 0.577$ is Euler's constant. This relation, derived for diffusive crossing of a symmetric parabolic 1D barrier with $\Delta G^* \gg k_BT$,[26] is commonly used for the analysis of experimentally measured biomolecular transition path times[3,11], where information about the shape of the potential is often unavailable. Figure 3c illustrates an important observation: In contrast to the transition rates, which depend exponentially on the activation free energy, as predicted by Kramers theory (Eq. 2, Figure 2c), the transition path times are much less sensitive to changes in barrier height. The observed dependence of $\langle\tau_{TP}\rangle$ on $\Delta G^*$ is also well described by the Szabo equation.

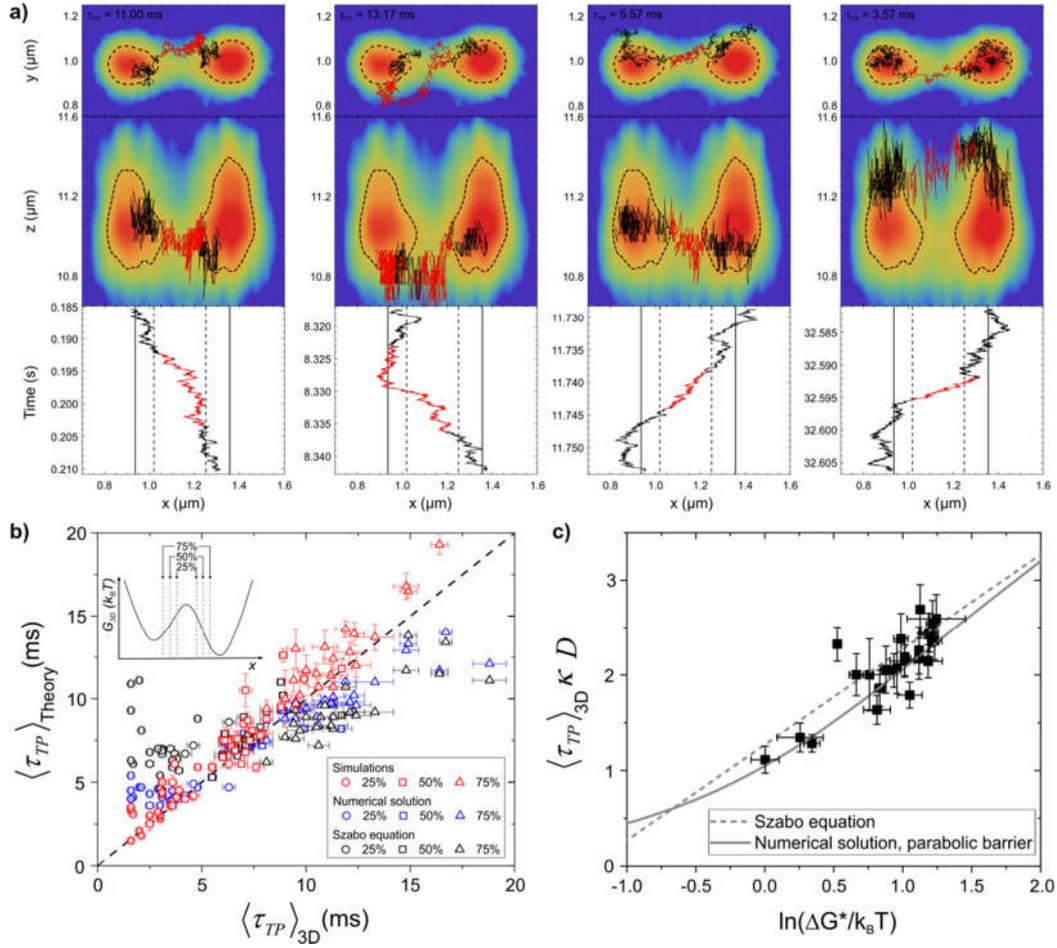

**Figure 3. Quantifying transition path times and dependence on barrier height. (a)** Examples of individual transition paths. The top panels show the transitions projected on the xy- and xz-plane, with the potential of mean force shown by the color scale as in Figure 1. The dashed line indicates the transition path boundary in the xy- and xz-plane, and the red segments of the trajectories indicate the transition paths identified in 3D. The bottom panels show the transition paths projected onto the x-axis as a function of time. The solid vertical lines indicate the centers of the traps, the dashed lines the transition path boundaries along x. **(b)** The experimentally determined average transition path times for each individual particle compared to the results obtained from 3D-Brownian dynamics simulations (red symbols), to the numerical solution of the 1D diffusion problem based on the measured potential of mean force (blue symbols), and to the results from theory using the Szabo equation assuming a high and parabolic barrier (black symbols) using different transition path boundaries (see inset and legend). For the theoretical values using the Szabo equation, the averages of $\Delta G^*$ in both directions of the transition were used. **(c)** Average transition path times for each individual particle using transition path boundaries at 50% (see inset in b) compared with the theoretical predictions according to A. Szabo's parabolic barrier approximation[3,4,26] (gray dashed line) and the numerical solution (solid gray line) using values for the curvature based on the experimentally determined potentials and the diffusion coefficient averaged over all measurements. Vertical error bars result from the uncertainties in the curvatures of the barrier, the diffusion coefficient, and the transition path time, which were obtained by bootstrapping. Horizontal error bars result from the difference in barrier heights for transitions starting in the left and right well.

The time resolution of our experiment allows us to quantify not only the average transition path time but also its distribution and the dependence on the barrier height for each individual particle (Figure 4a and Supplementary Figure 7). Consistent with the picture of diffusive barrier crossing, we observe peaked asymmetric distributions with exponential tails[25,27]. The distributions are compared with those predicted by three theoretical models: Brownian dynamics simulations using the 3D potentials, $G_{3D}(x,y,z)$, and diffusion coefficients from the experiments; the numerical solution for diffusive dynamics in the reconstructed 1D potential, $G_{1D}(x)$[25]; and the analytical approximation for the diffusive dynamics in a 1D parabolic barrier assuming a high barrier[27]. All three approaches are in remarkably good agreement with experiment for barriers with $\Delta G^* > 3\ k_BT$ (Figure 4a). For smaller barriers, the analytical approximation deviates from the exact solution and from the experimental data, whereas experiment, 3D simulation, and 1D numerical solution agree to within error down to very low barriers.

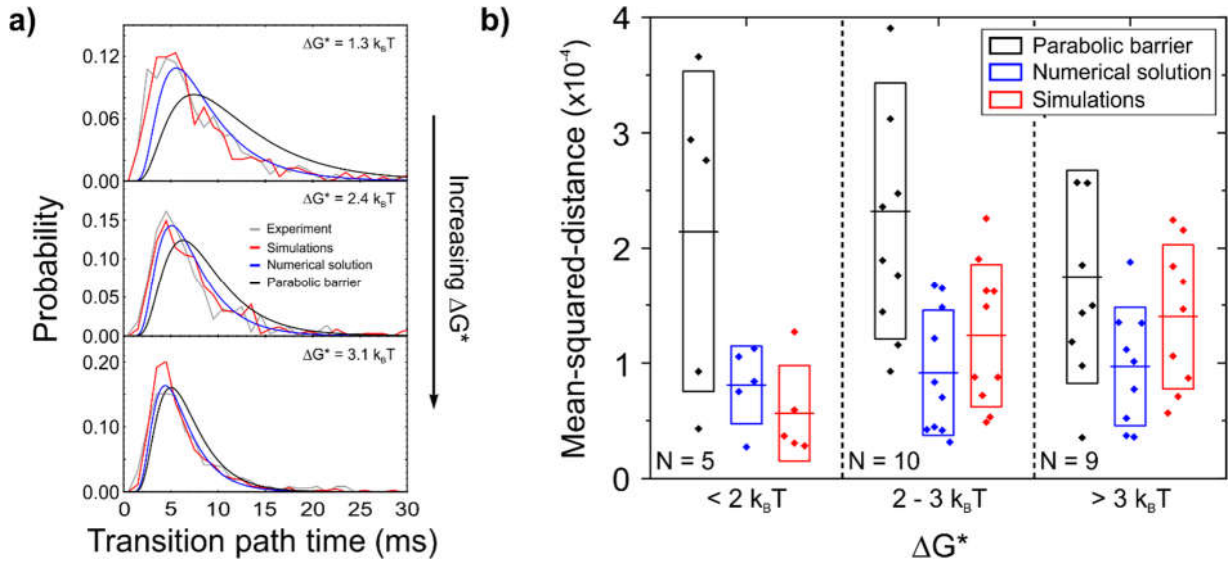

**Figure 4: Transition path time distributions for different barrier heights. (a)** Examples of transition path time distributions, each obtained from the measurement of an individual particle, with the barrier height given in each panel. Experimentally determined distributions (gray) are compared to 3D Brownian dynamics simulations (red), the numerical solution for 1D diffusion in the potential of mean force (blue), and the analytical approximation for a parabolic barrier (black). **(b)** Mean-squared-distance ($\chi^2$) between the experimental and theoretical distributions, divided into three groups of barrier heights: low ($\Delta G^* < 2\ k_BT$), intermediate ($2\ k_BT \leq \Delta G^* \leq 3\ k_BT$), and high ($\Delta G^* > 3\ k_BT$), where $\Delta G^*$ is the average of the barrier height in both directions of the transition. Symbols indicate individual measurements, the solid line is the average, and rectangles indicate the standard deviation of the mean-squared-distance. $N$ gives the number of distributions, i.e., the number of particles, contributing to each class.

The transition path time is an important property of transition paths, but much more information can be gleaned from the observed barrier crossing dynamics. For instance, most of the analyses of transition path times in single-molecule experiments have relied on the *a priori* assumption of diffusive dynamics. However, this assumption (which we expect to hold for a Brownian particle in solution but which is not necessarily justified for molecular reaction coordinates) can be tested directly using the observed trajectories. We proceeded in two different ways. First, we applied the recently proposed Markovianity test[28], which computes the conditional probability $P(x_1 \to x_2 | x)$ that the system at a point $x$ (with $x_1 < x < x_2$) is on a transition path leading from a boundary $x_1$ to boundary $x_2$. For a Markov process, this probability peaks at a value of 0.25, whereas memory effects reduce its maximum value. In our experiments, the observed barrier crossing dynamics indeed closely approach Markovianity (Figure 5c). The residual deviation from 0.25 is presumably caused by the rare misidentification of fast transition paths (Methods, Supplementary Figure 11).

Second, we applied a more general dynamic model to the observed trajectories, $x(t)$, in which the friction force is not simply proportional to the instantaneous velocity, $\dot{x}(t)$, but rather depends on the velocities in the past, $F_{fr} = -\int_{-\infty}^{t} \xi(t-t')\dot{x}(t')dt'$. The friction memory kernel, $\xi(t)$, can be reconstructed directly from the observed trajectories[29]. The Laplace transform of $\xi(t)$, $\hat{\xi}(s) = \int_{0}^{\infty} \xi(t)e^{-st}dt$, is close to a constant, $\hat{\xi}(s) = \xi_0$, in the experimentally accessible frequency range (Figure 5d), which means that the memory kernel is well approximated by a delta function, and the friction force is thus simply proportional to the instantaneous velocity, $F_{fr} = -\xi_0 \dot{x}$, providing direct evidence for diffusive dynamics. Moreover, the resulting estimate of the diffusion coefficient based on the Einstein relation, $D = k_B T / \xi_0$, agrees with the values from the mean square displacement and the position correlation function (Supplementary Figure 3). Finally, we also quantified the average shape of the transition paths[30-32] (Figure 5e) and the transition-path velocity profile[33], $v_{TP}(x)$ (Figure 5f), which characterizes the dynamics of barrier crossing in terms of the time derivative of the average shape. Consistent with the model of diffusive crossing of a parabolic barrier, these profiles are monotonic functions of the coordinate with a minimum at the barrier top[30,32,33].

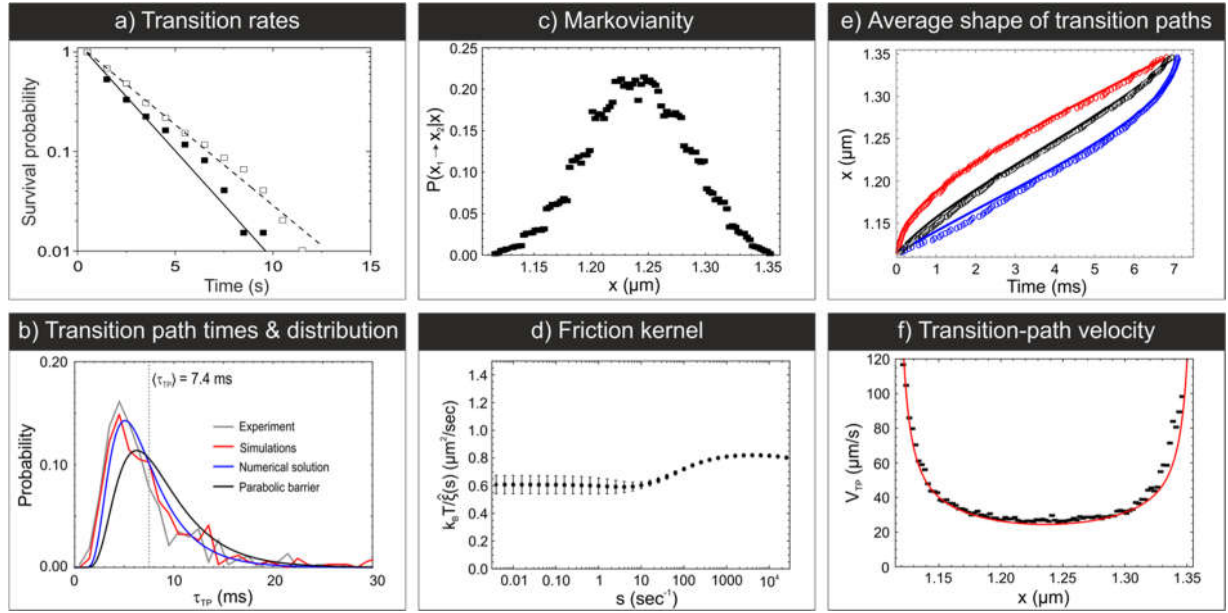

**Figure 5: Detailed information on transition paths.** The dynamics of individual nanoparticles in a bistable optical trap provides access to a wide range of information and direct comparison to theory. **(a)** The transition rates are available from the analysis of the dwell-time distributions. The high time resolution and the long-term stability of the measurements further enables quantification of **(b)** the full distributions of transition path times, **(c)** the Markovianity of the process[28], **(d)** the friction memory kernel[29] (Eq. 7), **(e)** the average shape of the transition paths (circles: experiment, lines: theory; red and blue: average first and last crossing times[30], black: average transition path shape[31], Eq. 14). Interestingly, the average transition path is nearly indistinguishable from the most probable transition path[34] estimated using the harmonic approximation[32] (Eq. 15, dashed gray line). **(f)** The transition-path velocity as defined by Berezhkovskii and Makarov (black symbols, Eq. 8) with the theoretical prediction[33] (red line, Eq. 9). See main text and Methods for details.

Our approach of high-speed tracking of nanoparticles in a bistable optical trap thus provides an opportunity to resolve transition paths of a diffusive barrier crossing process with great precision and to quantify their properties in unprecedented detail. Owing to the simplicity of the experimental system, a description of its dynamics in terms of diffusion in a potential of mean force is expected to be stringently applicable[5]. A key strength of the approach is that the parameters determining these dynamics – the 3D shape of the potential and the diffusion coefficient of the particle – can be extracted directly from the experimental data, essentially without simplifying assumptions. Moreover, the shape of the potential and especially the barrier height can be adjusted systematically. The results thus offer a stringent experimental benchmark for the theoretical concepts that have been and will be developed and increasingly used in the investigation of transition paths, especially in biomolecular processes[3,4,8-15]. Finally, experiments of this type and their analysis may help to address challenging open questions regarding biomolecular transition paths, such as the discrepancy between barrier heights using different types of analysis[10], or the observation of transition path velocity profiles with a maximum instead of a minimum[35].

**Acknowledgements**

We thank Karin Buholzer and Attila Szabo for valuable discussions and comments on the manuscript. This work was supported by the Swiss National Science Foundation (to B.S.), by the Forschungskredit of the University of Zurich (to N.Z.), by the Robert W. Welch Foundation (Grant F-1514 to D.E.M.) and by the US National Science Foundation (Grant CHE 1566001 to D.E.M).

**Author contributions**

B.S. and D.N. conceived the study. N.Z. and D.N. developed instrumentation and performed experiments. N.Z., D.N., R.S., and D.E.M. analyzed data. B.S., D.N., and D.E.M. supervised the project. All authors contributed to writing the paper.

**Competing interests**

The authors declare no competing interests.

## Methods

**Instrumentation**

Two optical traps were formed in an inverted microscope (Olympus IX-71) equipped with a water immersion objective (NA=1.2, 60x, PlanApo, Olympus) using a Nomarski prism and a linearly polarized laser beam from a continuous-wave solid state laser (λ = 485 nm, Sapphire 488-100 CDRH, Coherent) (Supplementary Figure 1). The relative intensities of the two traps were tuned by rotating a lambda-half plate positioned in the laser beam after a polarizer. Each of the two laser foci creates a stable three-dimensional potential well. The trap stiffness, and hence the depth of the potential wells, increases with increasing laser power. The distance between the two wells (~450 nm) is stably defined by the Nomarksi prism used. The sealed sample chamber, made from two glass cover slides separated by a polydimethylsiloxane spacer, contained a dilute solution of fused-silica microspheres (0.54 µm diameter, Bangs Laboratories) in water at room temperature. Using this configuration, we do not observe any drift in the position of a trapped microsphere over a duration of at least an hour (Supplementary Figure 2).

A single trapped particle was imaged with a high-speed CMOS camera (EoSens 4CXP CoaXPress, Mikrotron, Germany) using linearly polarized light from a 100-W mercury lamp using Köhler illumination. The camera was connected to a computer via a frame-grabber card (Active Silicon CoaXPress CXP-6 FireBird) and operated at 30,000 frames per second. Images were recorded directly to a high-speed SSD hard drive using the software StreamPix (NorPix inc., Canada). To achieve sufficiently high frame rates, the recorded region of interest was limited to 128 x 18 pixels, with a pixel size of 117 nm. To reduce the data throughput to a level compatible with the SSD write speed, 20 consecutive frames were read into the buffer of the frame grabber card and subsequently written on the SSD drive as a single frame. Before analyzing the movies, each frame was split into 20 individual frames using custom-written software in C++.

**Particle localization**

Before applying the localization routine, all frames were divided by a pre-recorded background image of the same area but without any particle present to reduce the influence of static camera noise. Subsequently, the coordinates of the center of the microsphere in the focal plane (xy) of the objective lens were found using an algorithm that determines the center of a radially symmetric pattern[22]. An accuracy of ~25 nm was achieved based on simulated particle images with comparable signal-to-noise as in the experiment. The coordinate in the direction of laser beam propagation (z) is extracted from the shape of the diffraction pattern of the particle image. For this purpose, we recorded calibration stacks using immobilized particles embedded in a 2% agarose gel (Supplementary Figures 8 and 9). The z-position of the particle was controlled via a piezo-actuated nanopositioner (PIFOC, Physik Instrumente, Germany) and adjusted in 50-nm increments over a range of 9 to 12.5 µm. Each calibration image was an average of 10,000 frames recorded at 30 kHz frame rate. After determining the center of symmetry, we obtained a radial intensity profile by angular integration. Finally, for each profile, the radial distance of the first extremum (minimum or maximum) from the center was determined (see Supplementary Figure 9). This value changes monotonically with the z-location of the particle and served for calibration. We repeated the calibration measurements on four different samples, measured on four different days and took the average. Subsequently, for each image in the experimental trajectories, the radial distance of the first extremum was determined analogously and the corresponding z-position determined according to the calibration curve.

**Extracting and characterizing the 3D potential**

From the extensive sampling of the particle's position (~$10^6$ to $10^8$ positions per recording), a detailed spatial probability density, $P_{3D}(r)$, can be obtained and related to the corresponding potential, $G(r)$, as a function of particle position according to

$$G_{3D}(r) = -k_B T \ln P_{3D}(r), \tag{1}$$

where *r* is the position vector of the particle, $k_B$ is the Boltzmann constant, and $T$ = 295 K is the temperature. To determine the potential, only data points with a z-position between 10.5 and 11.7 µm were used, where precise axial localization was possible (Supplementary Figures 8 and 9). Typically, less than 0.3% of the data points were outside this range and had to be discarded. Discarding these data points mainly influenced the resolution of the energy landscape at the top of the barrier, since the particle mainly travels outside the specified z-range when it is not stably trapped in one of the minima, and did not influence the transition paths and hence transition path times (see Supplementary Figure 10).

The shapes of the resulting free-energy barrier and the potential wells of the potential of mean force projected on the x-axis, $G_{1D}(x) = -k_B T \ln P_{1D}(x)$, can be approximated by a harmonic function, $G_{1D}(x) = 0.5 \kappa x^2$, where $\kappa$ is the curvature. To estimate the uncertainties in the curvatures, the potentials were fitted over three different ranges in energy (depending on the barrier height ranging from 0.25 to 3.0 $k_B T$), from which we determined the means and standard deviations of the curvatures. Most potentials were slightly asymmetric, resulting in different barrier heights for transitions starting from the left and right wells, respectively. For determining rate coefficients as a function of barrier height, dwell times for transitions in both directions were analyzed separately. For the analysis of mean transition path times and transition path time distributions, the barrier heights for transitions starting from both wells were averaged, and the resulting standard deviation used as the uncertainty in the height.

**Transition rates and comparison to Kramers theory**

Dwell time analysis was performed along the x-axis. To extract the dwell times and corresponding transition rates, the time traces were down-sampled from 30 kHz to 500 Hz by averaging the coordinates to reduce analysis time. To eliminate the influence of noise in the x-coordinate, which can lead to apparent transitions and shorten the resulting dwell times, we chose the threshold not half-way between the wells (as is used in standard thresholding methods), but at 75% of the distance between the wells relative to the starting well. To extract the transition rate coefficient, *k*, the survival probability, as a function of time *t*, was fitted with a single-exponential decay, $e^{-kt}$. These experimentally determined transition rates were compared with the transition rates for an overdamped Brownian particle in a potential as predicted by Kramers theory according to

$$k = k_0 e^{-\Delta G^\ddagger / k_B T}, \tag{2}$$

where $\Delta G^\ddagger$ is the barrier height in the potential of mean force projected on the x-axis, and $k_0$ the pre-exponential factor given by

$$k_0 = \frac{D}{2\pi k_B T} \sqrt{\kappa_b \kappa_w}, \tag{3}$$

where $D$ is the translational diffusion coefficient of the particle, and $\kappa_b$ and $\kappa_w$ are, respectively, the curvatures of the potential of mean force at the top of the barrier and at the bottom of the potential well. We note that the projected dynamics along more general one-dimensional coordinates may involve non-Markovian, memory effects – in such cases, Langer's multidimensional model can be mapped[36,37] onto the one-dimensional Grote-Hynes theory[38], which is a generalization of Kramers theory that accounts for memory. Due to the decoupling of the dynamics along $x$ near the barrier, however, these dynamics remain Markovian, and the Kramers description is adequate in our case. (This conclusion is further supported by the analysis presented in Figure 5, which shows that the dynamics is, indeed, Markovian and can be described by a one-dimensional Langevin equation.)

The mean-squared-displacement (MSD) of the particle in the x-coordinate was calculated according to (Supplementary Figure 3)

$$MSD = \frac{1}{N}\sum_{n=1}^{N}(x_n(t) - x_n(0))^2. \qquad (4)$$

$D$ was extracted from the MSD at times up to 165 µs, a range where the optical trap does not influence the motion of particle, using

$$MSD = 2Dt + c, \qquad (5)$$

where $c$ is the intercept of the MSD at $t = 0$, which is the result of the residual experimental uncertainty in the particle position. We found diffusion coefficients in the range of 0.73 – 0.89 µm$^2$ s$^{-1}$. The expected diffusion coefficient for a microsphere with a radius of 270 nm is 0.80 µm$^2$ s$^{-1}$ according to the Stokes-Einstein relation, corresponding to the expected particle-to-particle variation in radius of ±10 - 15% specified by the supplier. Diffusion coefficients from the MSD at short times agreed with the diffusion coefficients obtained from the diffusive component of the position correlation function (Figure 1c) according to $D = k_BT/\tau_r k$, where $\tau_r$ is the observed relaxation time and $k$ the spring constant of the potential well.

**Identifying transition paths**
Transition path analysis was performed using a custom-written routine (Wolfram Mathematica 11). The boundaries for the start and end of transition paths were chosen half way between the energy top of the barrier and minimum of the well based on an interpolation of the measured 3D energy landscape (using the Mathematica function *ListInterpolation*, with method spline and order 3). The start of a transition trajectory was defined as the first data point with an energy greater than the initial boundary, while the end of the transition trajectory was the first data point with an energy less than the final boundary. The transition trajectory was only identified as a transition path if the start and end points were in different potential wells. From the set of transition paths obtained for each individual particle, only the transitions for which more than 80% of the data points had a z-position within 10.5 and 11.7 µm were used to determine averages and distributions of transition path times. This procedure eliminated artificial rapid transitions (less than ~1 ms, caused by a lack of contrast of the images outside this range and resulting in unreliable particle positions), whereas the rest of the transition path distributions was virtually unaffected (Supplementary Figures 10 and 11).

To validate our experimental methodology of identifying transition paths in the experimental data, we performed 3D-Brownian dynamics simulations of a particle on the interpolated energy landscapes with the experimentally determined diffusion coefficients and generated time traces of similar length as in the experiments. The simulations were performed with time-steps equivalent to a frame rate of 250 kHz to also be able to test whether the experimental frame rate of 30 kHz was sufficient for accurately identifying transition paths. Moreover, normal distributed noise was added to the simulated time traces; its standard deviation (7 nm) was chosen so that the intercept in the resulting mean-squared-displacement was similar to the one found experimentally (see Eq. 5). The resulting time traces were analyzed in the same way as the experimental data and compared to the experimental results in terms of barrier heights, transition rates, average transition path times, and transition path time distributions (Supplementary Figure 12).

**Estimating the friction memory kernel $\xi(t)$ from experimental trajectories**

Assuming that the dynamics are governed by an overdamped generalized Langevin equation of the form

$$0 = -G'_{1D}(x) - \int_{-\infty}^{t} \xi(t-t')\dot{x}(t')dt' + \zeta(t), \qquad (6)$$

where $\zeta(t)$ is a Gaussian-distributed random force with zero mean that satisfies the fluctuation-dissipation theorem, $\langle \zeta(t)\zeta(t')\rangle = k_B T \xi(t-t')$, we estimated the friction memory kernel using the exact formula[29]:

$$\hat{\xi}(s) = \hat{C}_{fx}(s)\left[s\hat{C}_{xx}(s) - C_{xx}(0)\right]^{-1}. \qquad (7)$$

Here, $\hat{\xi}(s) = \int_0^{\infty} \xi(t)e^{-st}dt$ is the Laplace transform of the memory kernel, and $\hat{C}_{xx}(t)$ and $\hat{C}_{fx}(t)$ are the Laplace transforms of the position-position and force-position correlation functions, $C_{xx}(t) = \langle x(t)x(0)\rangle$ and $C_{fx}(t) = \langle -G'_{1D}[x(t)]x(0)\rangle$, respectively. The memory kernel is reported in Laplace space in Figure 5.

**Estimating the average shape and velocity of transition paths.**

The effective velocity of transition paths at point $x$, $v_{TP}(x)$, is defined[33] as the ratio of the length of a short interval, $\Delta x$, centered at $x$ and the mean cumulative time spent by transition paths in this interval. It was estimated using the formula[33]

$$\frac{1}{v_{TP}(x)} = \langle \tau_{TP}\rangle P(x|TP), \qquad (8)$$

where $P(x|TP)$ is the probability density of visiting point $x$, given that the system is on a transition path. The velocity profile was computed directly using Eq. 8. It was also compared to the exact analytical formula valid in the case of diffusive dynamics with a constant diffusion coefficient, $D$,[33]

$$\frac{1}{v_{TP}(x)} = \left(\int_{x_1}^{x_2}\frac{dx'}{DP_{eq}(x')}\right)\Phi_1(x)\Phi_2(x)P_{eq}(x), \qquad (9)$$

where $P_{eq}(x)$ is the equilibrium distribution of the reaction coordinate, $x$, and $\Phi_{1(2)}(x) = 1 - \Phi_{2(1)}(x)$ is the splitting probability, i.e., the probability, having started from $x$, to reach the boundary $x_1$ ($x_2$) before the boundary $x_2$ ($x_1$). The splitting probability is given by

$$\Phi_1(x) = 1 - \Phi_2(x) = \int_{x_1}^{x} P_{eq}^{-1}(x')dx' \bigg/ \int_{x_1}^{x_2} P_{eq}^{-1}(x')dx'. \tag{10}$$

We note that by taking the time derivative of the average transition path, $\tilde{v}_{TP} = dx_{TP}(t)/dt$, one obtains an alternative definition of the transition path velocity.[32]

We have also estimated the average shape of the transition path, $x_{TP}(t)$, which is defined[30] through the inversion of the dependence of the average times, $\langle t_{first}(x_{TP})\rangle$ and $\langle t_{last}(x_{TP})\rangle$, that a transition path from $x_1$ to $x_2$ takes to cross a point $x_{TP}$ for the first time and for the last time before arriving at the boundary $x_2$. Since this definition leads to a path that is not time-reversible, we have also estimated the symmetrized average path shape [30,32] defined via inversion of the coordinate dependence of the symmetrized average time

$$\langle t_{sym}(x)\rangle = \frac{\langle t_{first}(x)\rangle + \langle t_{last}(x)\rangle}{2}. \tag{11}$$

For diffusive dynamics, exact analytical expressions exist for these shapes, which are compared with the experimental results in Fig. 5f. Specifically,

$$\langle t_{first}(x)\rangle = \langle \tau_{TP}(x_1 \leftrightarrow x)\rangle \tag{12}$$

and

$$\langle t_{last}(x)\rangle = \langle \tau_{TP}(x_1 \leftrightarrow x_2)\rangle - \langle \tau_{TP}(x \leftrightarrow x_2)\rangle, \tag{13}$$

where $\langle \tau_{TP}(x_1 \leftrightarrow x_2)\rangle$ is the mean transition path time between the boundaries $x_1$ and $x_2$ (and similar definitions, with boundaries changed accordingly, are used in Eqs. 12, 13), given by the analytical expression[4]:

$$\langle \tau_{TP}(x_1 \leftrightarrow x_2)\rangle = \left(\int_{x_1}^{x_2} \frac{dx'}{DP_{eq}(x')}\right)\int_{x_1}^{x_2} \Phi_1(x)\Phi_2(x)P_{eq}(x)dx. \tag{14}$$

The most probable transition path from x = -L to L using the harmonic approximation was calculated according to[32]:

$$\langle x(t \mid \tau_{TP})\rangle = \frac{L\sinh\left(\kappa D/k_B T\left(t - \tau_{TP}/2\right)\right)}{\sinh\left(\kappa D/k_B T\left(\tau_{TP}/2\right)\right)}. \tag{15}$$

Supplementary Information for:

# Transition path dynamics of a nanoparticle in a bistable optical trap

Niels Zijlstra[1,4], Daniel Nettels[1], Rohit Satija[2], Dmitrii E. Makarov[2], Benjamin Schuler[1,3]

[1]Department of Biochemistry, University of Zurich, 8057 Zurich, Switzerland.

[2]Department of Chemistry and Oden Institute for Computational Engineering and Sciences, University of Texas at Austin, Austin, Texas 78712, USA.

[3]Department of Physics, University of Zurich, 8057 Zurich, Switzerland.

[4]Current address: Physical and Synthetic Biology, Faculty of Biology, Ludwig-Maximilians-Universität München, Großhaderner­strasse 2-4, 82152 Planegg-Martinsried, Germany

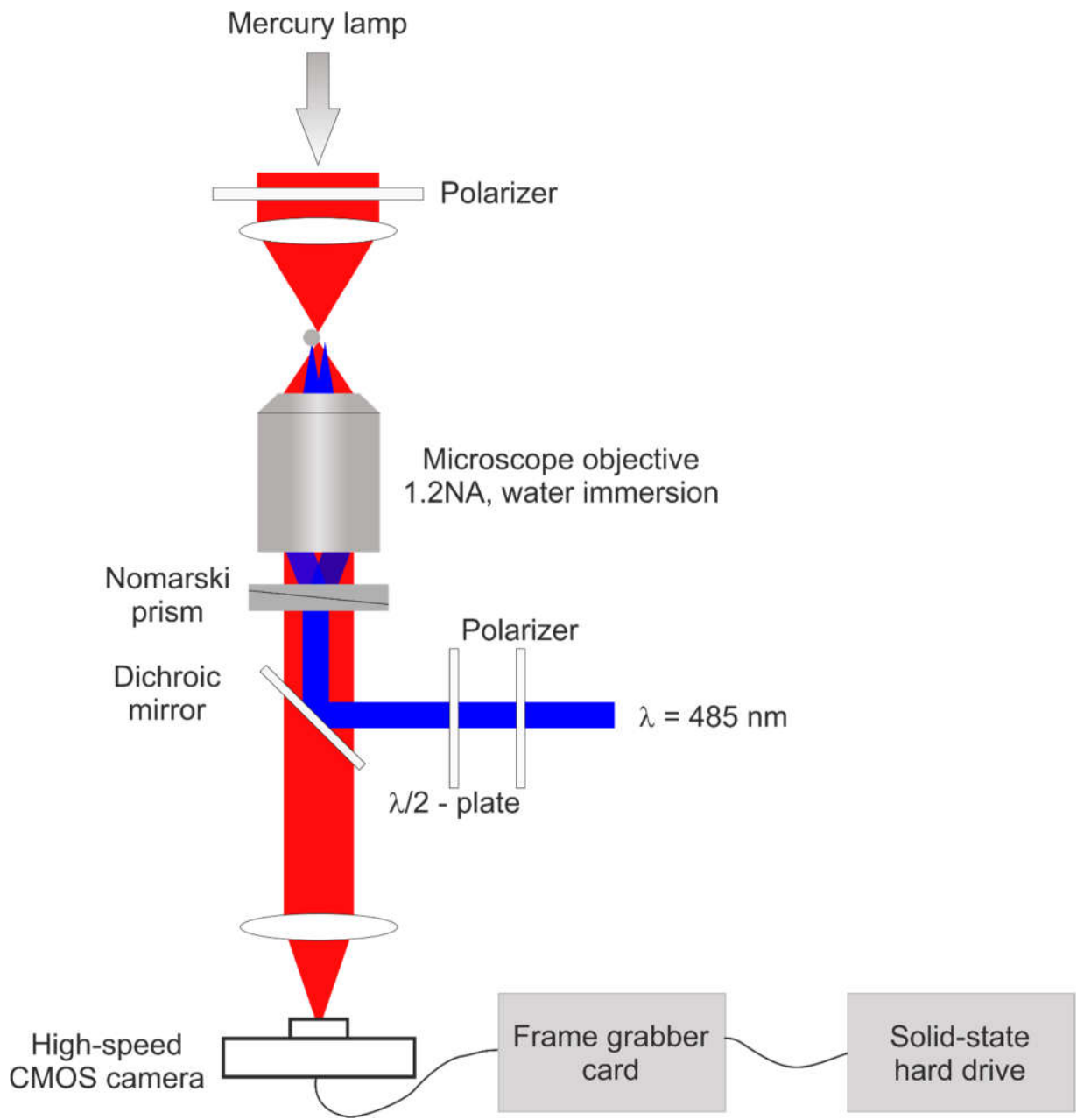

**Figure S1: Experimental setup.** A linearly polarized laser beam is focused by a microscope objective after passing through a Nomarski prism, which splits the light into two orthogonally polarized beams, leading to two optical traps with a stable displacement of ~400 nm in the focal plane of the high-aperture microscope objective. The relative stiffnesses of the traps are adjustable by rotating the lambda-half plate. The particle position is imaged onto a high-speed CMOS camera using Köhler illumination with the light of a mercury vapor lamp. A fast frame grabber card stores the images on a solid-state drive.

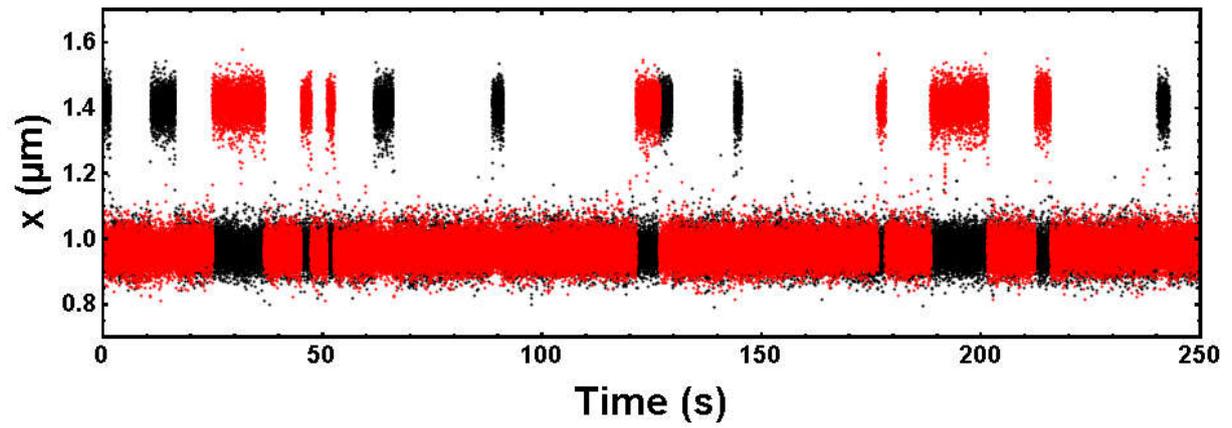

**Figure S2: Long-term stability of the experimental setup** demonstrated by the superposition of the first and last 250 seconds (black and red) of a 3300 s long time trace of the bead's x-position. No instrument drift is visible over this time span.

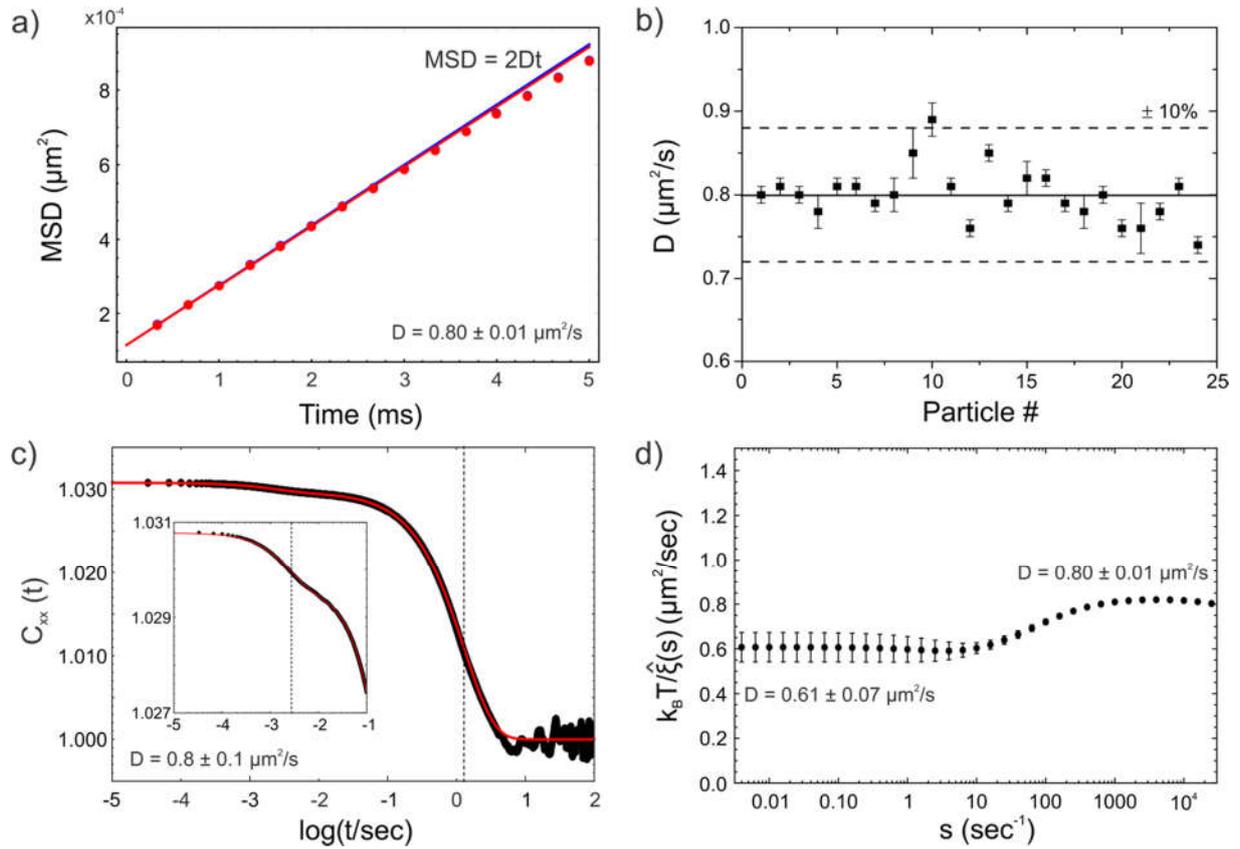

**Figure S3: Quantifying translational diffusion coefficients. a)** For each particle, the diffusion coefficient is determined from its mean-squared displacement (MSD) in the potential wells at short times (see Methods). The red and blue data points are examples obtained from the particle residing in the left and right well, respectively. The offset in the MSD at time zero is a result of the uncorrelated uncertainty in the particle's position due to noise in the images. **b)** The diffusion coefficients for all particles were within the 10-15% deviation from the mean value, which matches the diffusion coefficient expected for a sphere with a radius of 270 nm, as determined via the Stokes-Einstein equation, in accord with the supplier's specifications. **c)** The position autocorrelation function, $C_{xx}$, of a particle in the bistable optical trap shows components corresponding to the diffusive dynamics within each trap (relaxation time of 2.7 ± 0.1 ms in this example, corresponding to a diffusion coefficient of 0.8 ± 0.1 µm²/s, see inset with zoom) and to the kinetics of transitions between them (relaxation time of 1.3 ± 0.1 s in this example, in agreement with the inverse sum of the transition rate coefficients, 1.1 ± 0.3 s). **d)** The estimate of the diffusion coefficient based on the friction coefficient from the memory kernel at high frequency, where the error is smallest, and the Einstein relation, $D = k_B T / \xi_0$, yields 0.80 ± 0.01 µm²/s, in agreement with the value from the mean square displacement (0.80 ± 0.01 µm²/s) and the position correlation function (0.9 ± 0.1 µm²/s).

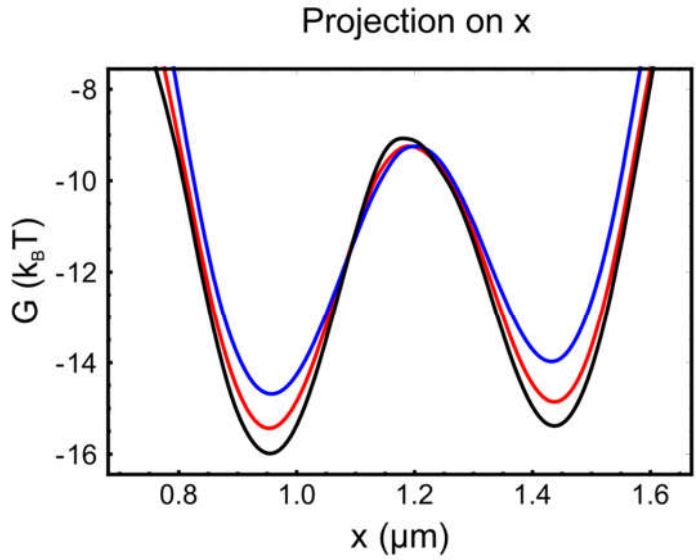
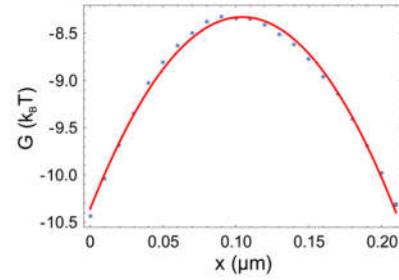
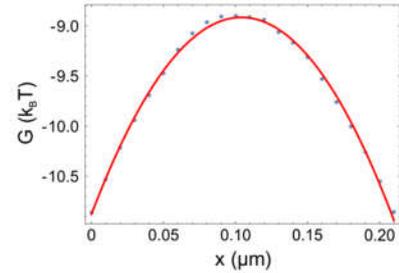
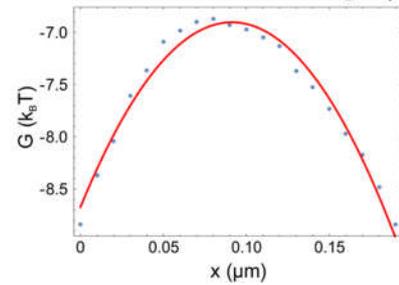

**Figure S4: Particle-to-particle variability leads to measurable differences in the trapping potential.** Potentials of mean force extracted for three particles with slightly different radii but at identical laser power. For each particle, the shape of the potential of mean force is measurably different, with pronounced differences in well depth/barrier height (left side) and slight differences in the curvature at the top of the barrier (right side).

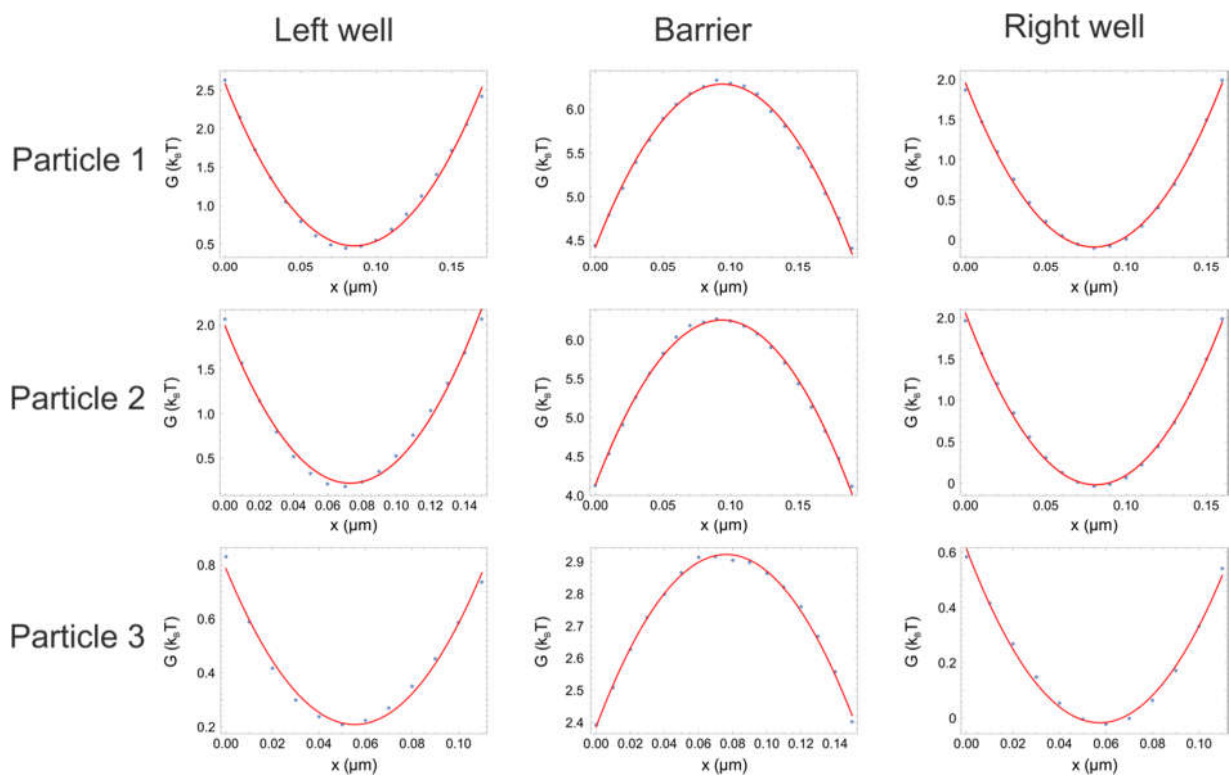

**Figure S5: Analysis of curvatures.** Examples of curvatures from fits of the energy barriers and wells of the potentials of mean force of three different particles at three different laser powers. Each curve was fitted over three different ranges of energies between 0.5 and 3 $k_BT$ relative to the top of the barrier or bottom of the well. The means and standard deviations of the extracted curvatures were used for the further analysis in terms of Kramers theory or barrier crossing.

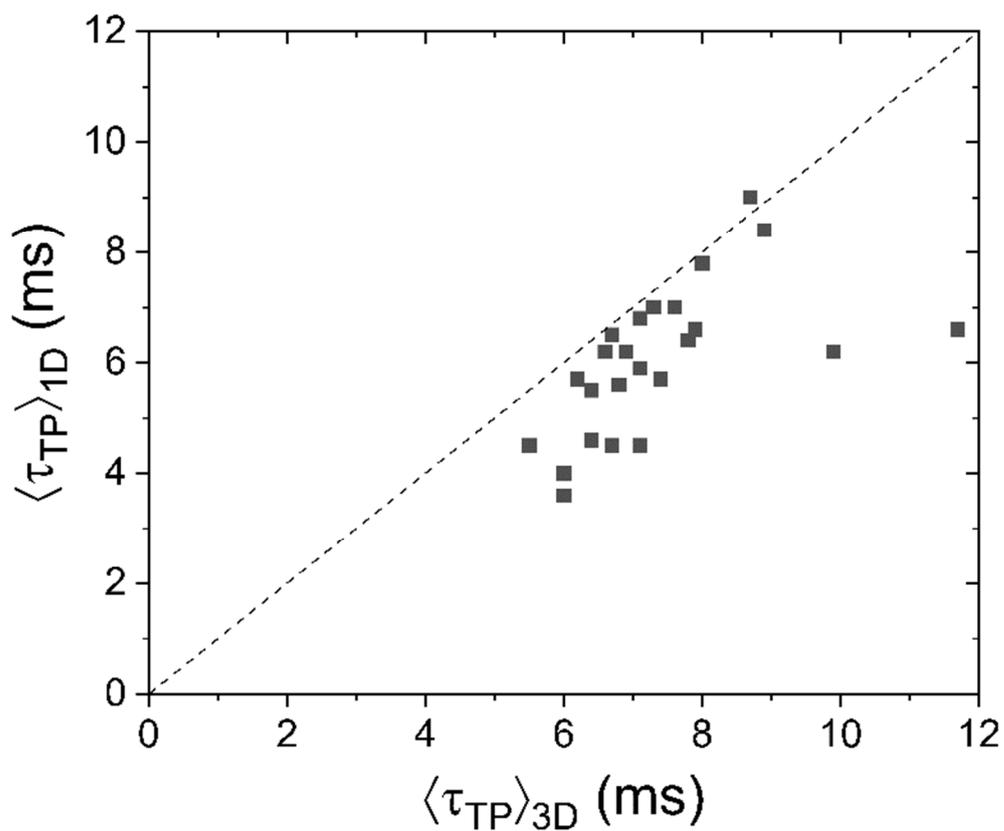

**Figure S6: Comparison of mean transition path times obtained from 1D- and 3D-analysises of the experimental data.** Transitions path times extracted from the 1D trajectories along the x-coordinate are on average ~15% shorter than the ones extracted from the 3D trajectories owing to the different choice of transition path boundaries in the two types of analysis.

# $\Delta G^* = < 2\ k_B T$

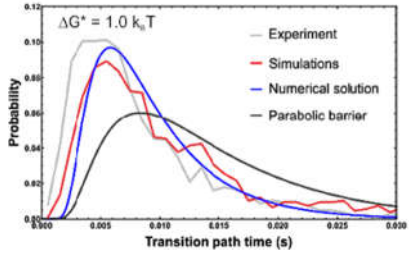
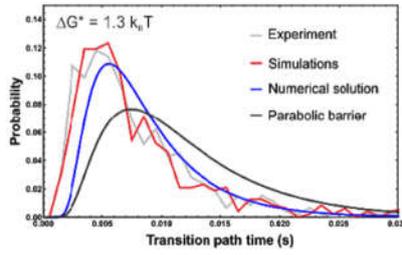
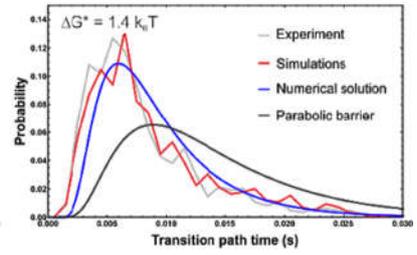
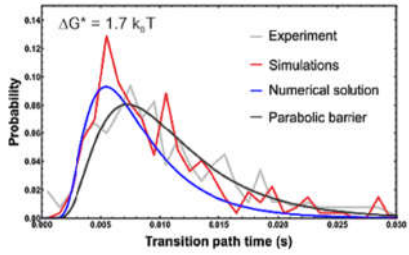
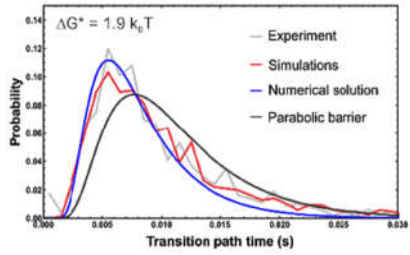

# $\Delta G^* = 2 - 3\ k_B T$

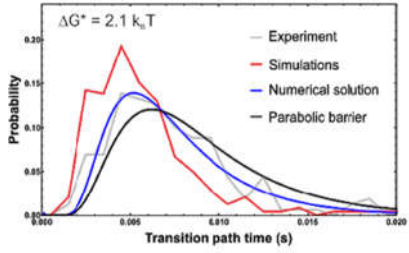
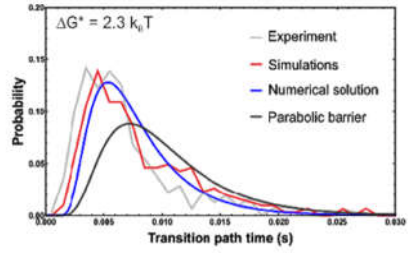
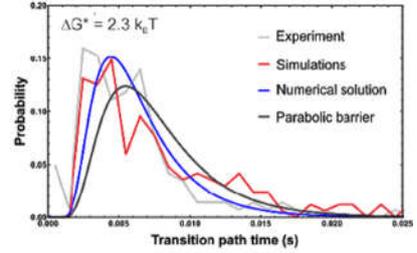
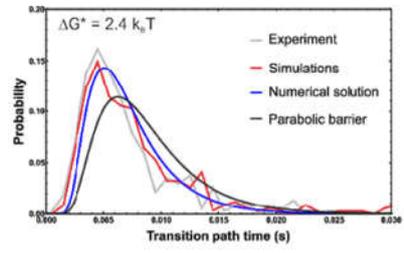
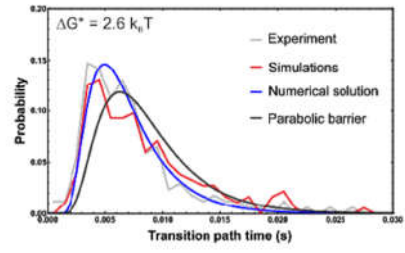
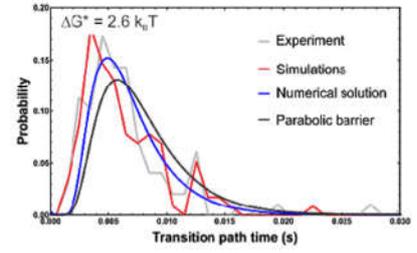
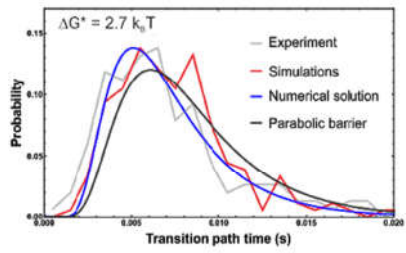
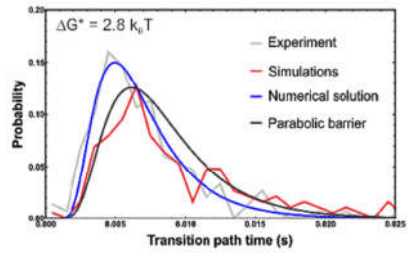
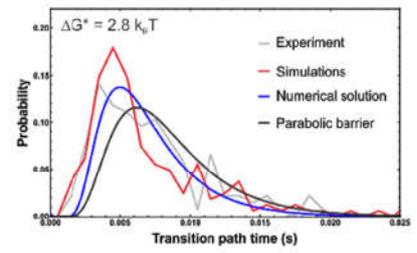
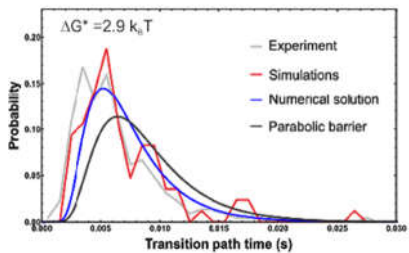

$\Delta G^* = > 3\ k_BT$

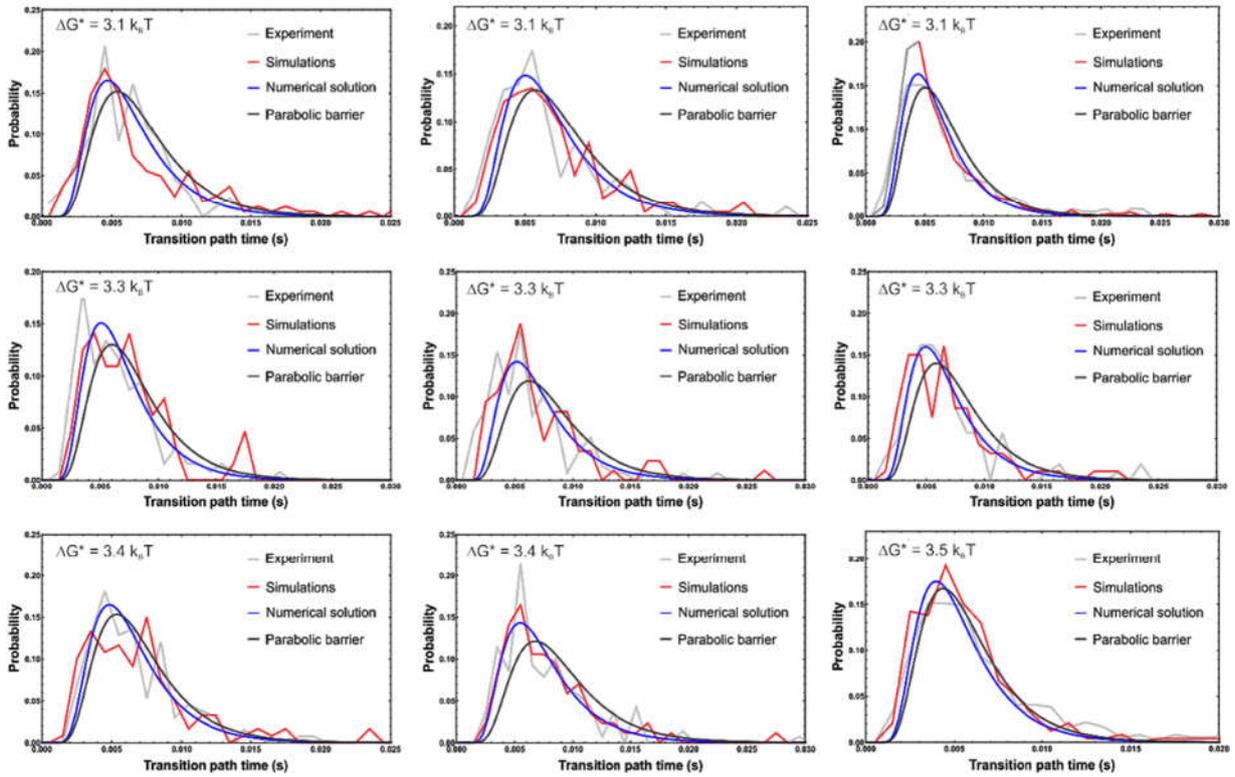

**Figure S7: Transition path time distributions.** Distributions of transition path times obtained from the 3D trajectories of each individual particle sorted by Δ*G*\*, where ΔG\* is the average of the two barrier heights relative to the left and right well. Each experimentally obtained distribution (gray) is compared to the distribution obtained from Brownian dynamics simulations (red), the numerical solution of the diffusion problem (blue), and the parabolic barrier approach (black, Szabo equation). Since the durations of the simulated trajectories (and therefore the total number of transitions therein) were chosen to be similar to the experimental ones, the noise in the distributions from the simulated and experimental data is also similar.

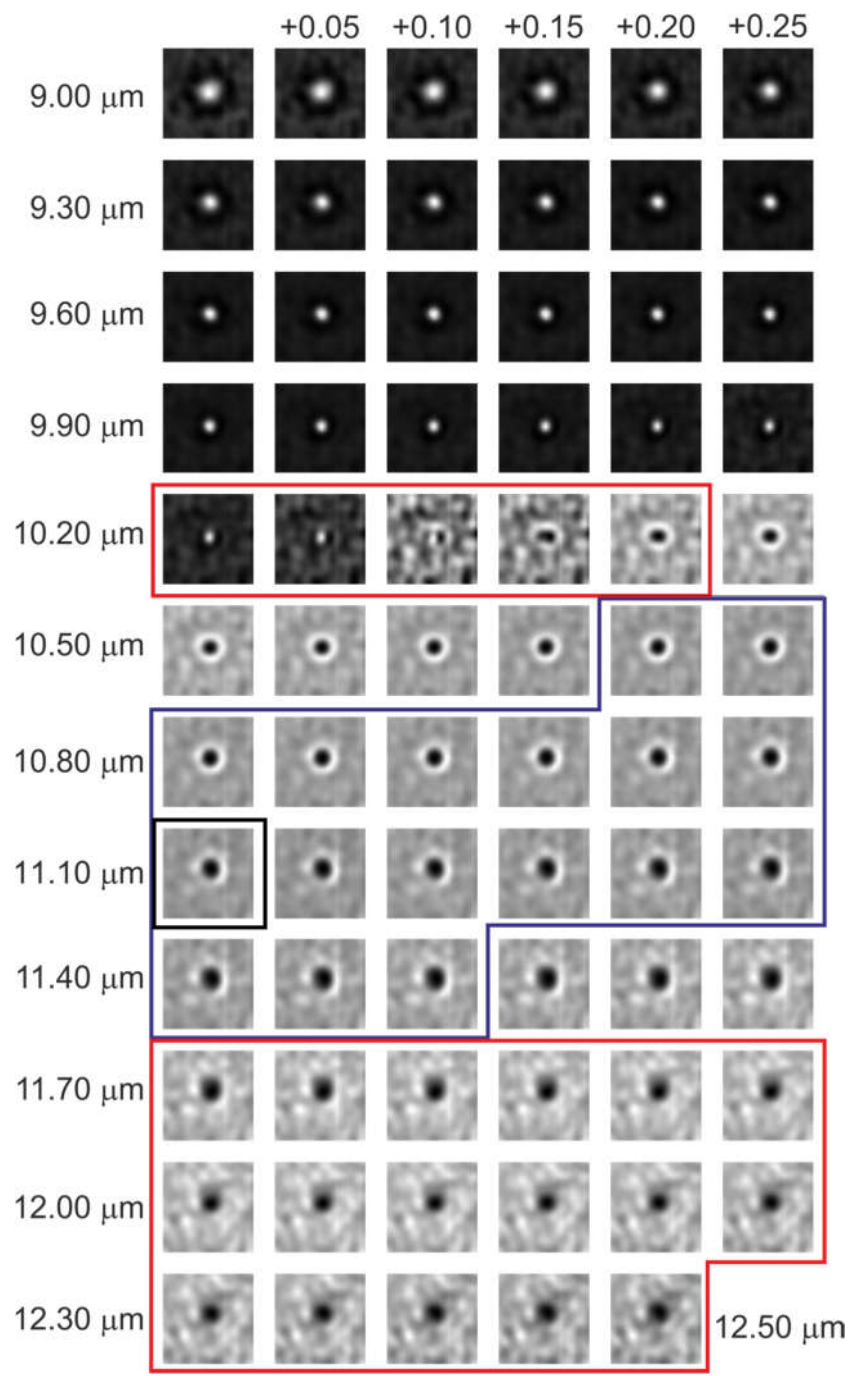

**Figure S8: Calibration of z position.** A calibration stack of images is generated from particles embedded and thus immobilized in a 2% agarose gel. Each image in the stack is the average of 10,000 images recorded at 30-kHz frame rate. The z-position was adjusted in 50-nm increments. The blue line indicates the most occupied range of z positions when the particles are trapped; the image framed in black shows the typical center position. The red lines indicate where determining the position of the particle is problematic due to a lack of contrast. The number on the left indicates the z position for the left-most image.

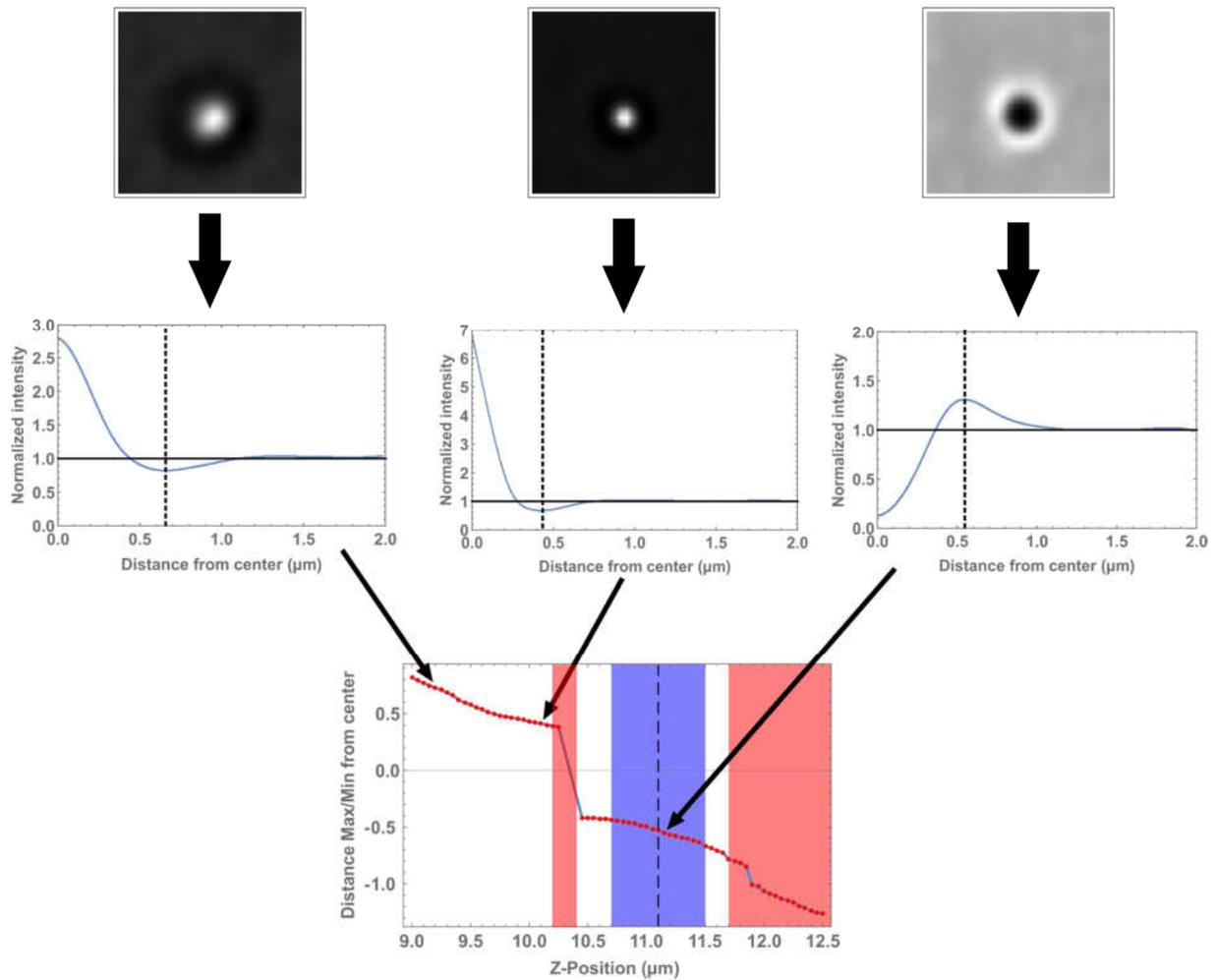

**Figure S9 Determining the particle position in z.** For each image from the calibration stack (see Supplementary Figure S8 for a full stack), the center of the particle is determined using the radial symmetry of the diffraction pattern. We obtain a radial intensity profile, $I(r)$, by angular integration and normalizing to the signal far from the bead, such that $I(r \gg 0) = 1$. Next, the distance, $r_z$, between the center of symmetry and the first minimum or maximum is determined. Finally, we construct a calibration curve by taking $+r_z$ if $I(r = 0) > 1$, and $-r_z$ if $I(0) \leq 1$. The calibration curve shown here is the average of four independent calibration stack recordings. The blue region indicates the most occupied z-range of a trapped particle, with the black dashed line being the average location of the particle. The red areas indicate ranges where the localization precision of the bead is reduced due to a lack of contrast (see Figure S8).

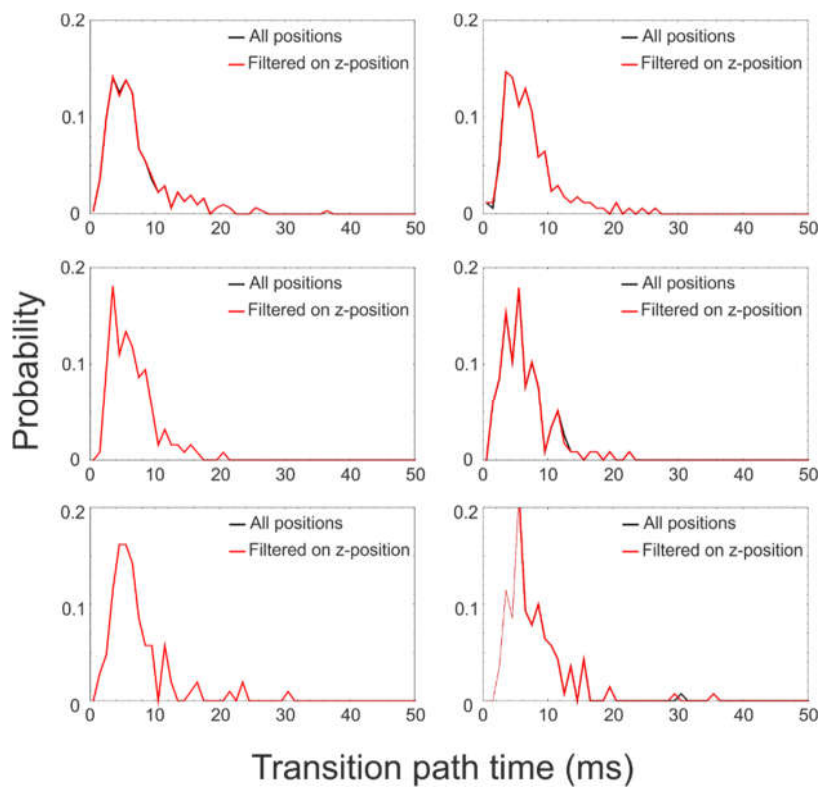

**Figure S10: Robustness of the transition-path-time determination on the range of z-positions used for reconstructing the energy landscape.** Comparison of the transition path time distributions based on all bead positions (black) with those using only the positions with z-values between 10.5 and 11.7 µm (red) shows no significant differences. For this comparison, simulated 3D-trajectories (Brownian motion on six of the experimentally determined energy landscapes) were analyzed.

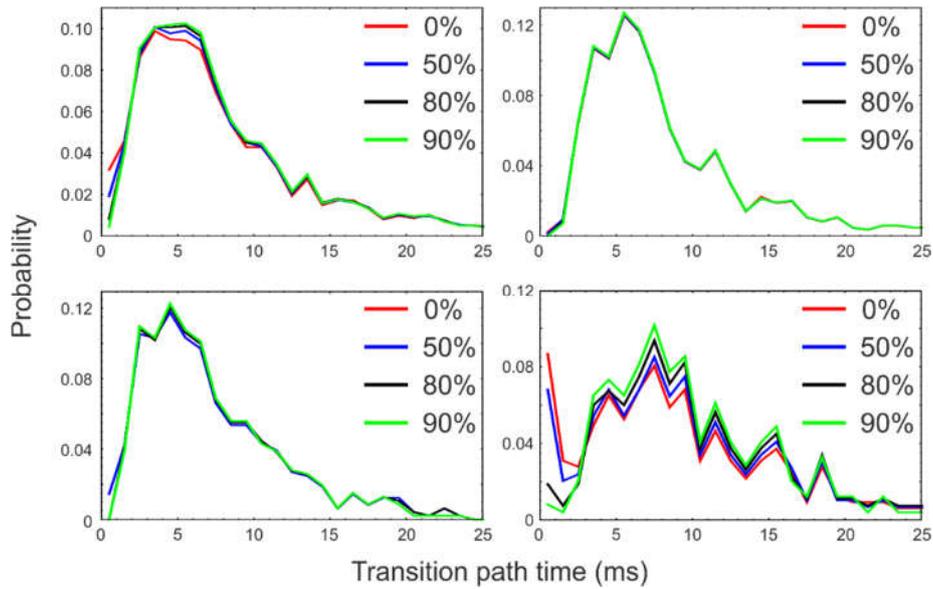

**Figure S11: Measured transition path time distributions are robust to the range of z positions used.** Effect of including only the transition paths for which more than a given percentage of the data points have a z-position between 10.5 and 11.7 µm. As the distributions show, an increasing percentage mainly excludes very fast apparent transitions with a duration of less than 1 ms owing to a lack of precise position information, most of which are eliminated with a threshold of 80%. Increasing the percentage from 80% to 90% resulted in discarding some longer transitions for which we can still determine the transition path time accurately due to the large number of frames within the transitions.

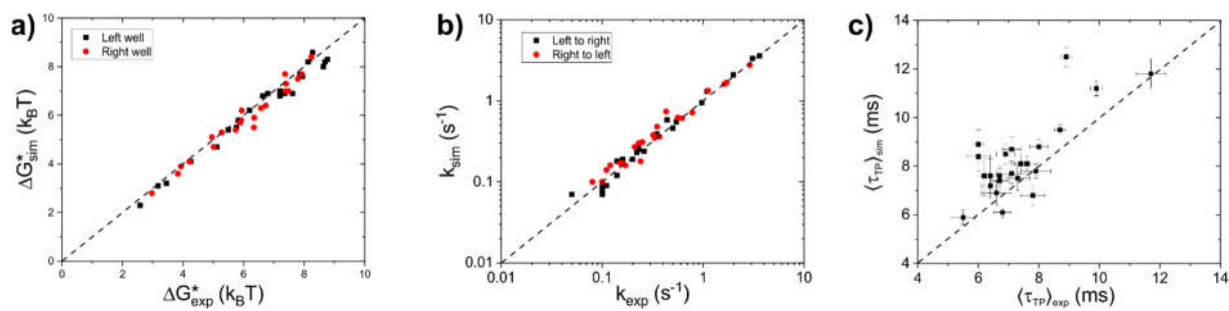

**Figure S12: Comparison of 3D-Brownian dynamics simulations and experimental data.** Experimentally obtained energy landscapes and diffusion coefficients were used for the simulations, which were then analyzed in the same way as the experimental data. **a)** The barrier heights extracted from experiment and simulation agree well. Red and black data points represent the barrier heights with respect to the left and right wells, respectively. The dashed curve marks the identity line. **b)** The transition rates between the wells, in the same representation as in **a**. **c)** The average transition path times obtained from experimental and simulated data. The error bars were obtained from bootstrapping.